\documentclass[11pt,aps,tightenlines]{revtex4}

\usepackage{amsmath}
\usepackage{amssymb}
\usepackage{bbm}
\usepackage{mathtools}
\usepackage[normalem]{ulem}
\usepackage{dsfont}
\usepackage{mathrsfs}
\usepackage{cancel}

\usepackage{braket}

\usepackage[utf8]{inputenc}
\usepackage[T1]{fontenc}

\def\la{\langle}
\def\ra{\rangle}

\def\i{{\bf i}}
\def\j{{\bf j}}

\def\B#1{\!\left(#1\right)}

\def\be{\begin{equation}}
\def\ee{\end{equation}}

\def\bee{\begin{equation*}}
\def\eee{\end{equation*}}

\def\ba{\begin{equation}\begin{aligned}}
\def\ea{\end{aligned}\end{equation}}

\def\Var{{\rm Var}}
\def\dVar{\partial_\eta\!\Var}
\def\dVartilde{\partial_{\tilde\eta}\!\Var}
\def\EIsing{{\cal E}_\text{Ising}}
\def\dwaE{\partial^2_g\EIsing}

\def\pref{0.65}

\def\for{\ \text{for} \ }
\def\kT{k_BT/U} 

\makeatletter
\newcommand{\pushright}[1]{\ifmeasuring@#1\else\omit\hfill$\displaystyle#1$\fi\ignorespaces}
\newcommand{\pushleft}[1]{\ifmeasuring@#1\else\omit$\displaystyle#1$\hfill\fi\ignorespaces}
\makeatother

\usepackage[ddmmyyyy]{datetime}
\makeatletter
\def\Dated@name{}
\makeatother

\begin{document}

\title{Critical points of the three-dimensional Bose-Hubbard model from on-site atom number fluctuations}
\author{Oskar A. Prośniak, Mateusz Łącki, and Bogdan Damski}
\affiliation{Jagiellonian University, Institute of Physics, {\L}ojasiewicza 11, 30-348 Krak\'ow, Poland}
\begin{abstract}

We discuss how  positions of  critical points of the three-dimensional
Bose-Hubbard model can be accurately obtained from variance of the on-site
atom number operator, which can be experimentally measured. The idea that we explore is that the derivative of the
variance, with respect to the parameter driving the transition, has a pronounced
maximum  close to  critical points. 
We show that  Quantum Monte Carlo studies of this maximum 
lead to  precise determination of  critical points for the superfluid-Mott
insulator transition in systems with mean
number of atoms per lattice site equal to one, two, and three. 
We also extract from such data the correlation-length 
critical exponent  through the finite-size scaling analysis
and discuss  how  the  derivative of the  variance  can
be reliably computed from  numerical data for the variance.
The  same conclusions apply to the derivative of the 
nearest-neighbor correlation function, which can be obtained from 
routinely measured  time-of-flight images.

\end{abstract}
\date{\today}
\maketitle

The field of quantum phase transitions is significant for  both fundamental and
practical reasons \cite{Sachdev,SachdevToday,ContinentinoBook}. 
On the fundamental side, it provides insights into 
rich physics of the most complicated many-body quantum systems whose description 
challenges the most advanced theoretical and numerical  studies.
On the practical side,
understanding of quantum phase transitions is critical for explanation
of  properties  of numerous  materials of potential technological importance
 such as high-temperature superconductors. 
One of the key obstacles to progress in the field of quantum phase transitions 
is our inability to efficiently solve the  models describing
strongly-correlated systems.   
 
The recent progress in cold atomic physics suggests that it could  be possible to
approach this problem  from another angle. Namely, one can use cold atomic
setups as hardware for simulation of condensed matter models. 
The idea to use an easy-to-control quantum system
to simulate another quantum system that is hard-to-study in conventional
setups dates back to  Feynman \cite{Feynman}.
Its proof of principle experimental demonstration was achieved about two decades ago  
in the cold atom experiment \cite{GreinerNature2002}. 
Ever since this publication, the field of quantum simulation has attracted  a
lot of attention  by  providing  both a fertile ground for interdisciplinary research
and a plausible promise of delivering  flexible experimental hardware for solving 
various   challenging models. 

\begin{figure}[t]
\includegraphics[width=\pref\textwidth,clip=true]{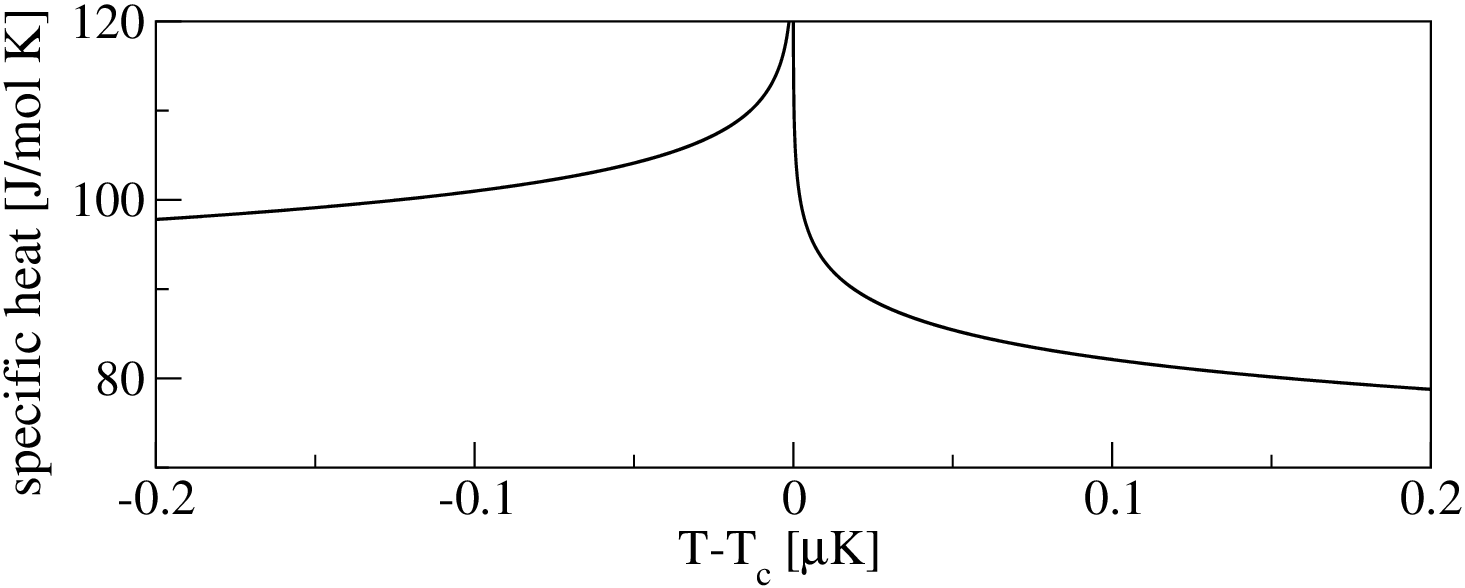}
\caption{Specific heat of liquid   $^4$He near the
lambda transition \cite{LipaPRB2003}. 
The curve is given by $A|1-T/T_c|^{-\alpha}+B$, where 
$\alpha\approx-0.0127$, $B\approx460$  $\text{J}/(\text{mol}\cdot\text{K})$, and $T_c\approx2.17$ $\text{K}$. 
The $A$ coefficient for $T<T_c$ and $T>T_c$ is approximately
given by $-447$ $\text{J}/(\text{mol}\cdot\text{K})$  and $-471$ $\text{J}/(\text{mol}\cdot\text{K})$, respectively.
Its asymmetry produces the lambda shape of the specific heat curve. 
Specific heat is not divergent at the critical point because the exponent
$\alpha<0$.
Measurements of specific heat as large as
$120$ $\text{J}/(\text{mol}\cdot\text{K})$ were reported.
A similar shape is observed in our results for $\dVar$.
}
\label{fig1}
\end{figure}

At the center of these studies is the implicit assumption that such
models can be  {\it quantitatively} studied in cold atom simulators.
Despite   fascinating experimental efforts that have been reported so far 
\cite{LewensteinAdv,BlochRMP2008}, 
the current characterization of quantum phase transitions in
best  cold atom
simulators is nowhere near the level of accuracy that is found in top condensed matter 
experiments. 

To justify this statement and to place the research discussed  in this paper  in a
broader context, we compare measurements of the
lambda transition in liquid   $^4$He to measurements of  the superfluid--Mott insulator
transition in the two-dimensional (2D) cold atom cloud. Such a comparison is 
meaningful because the two systems are expected  to   belong to the same
universality class, the one of the 3D $XY$ model.

The experimental studies of the lambda transition in $^4$He have started about a century ago 
 and continue to this date. The good reason for this long-term 
interest is that  they  provide  the most stringent tests of the renormalization-group 
(RG) theory,
which is the cornerstone of our understanding of phase transitions.
The RG  predictions are best established very close to the critical point
\cite{Cardy2}, 
where it is difficult to do measurements.
One of the key current  problems in accurate testing of the RG theory at the lambda transition is that 
gravity broadens the transition (the transition temperature depends on pressure, which 
varies across the sample due to the gravitational field). 
To remove this obstacle, the measurements  were 
taken in a zero-gravity environment in a Space Shuttle. 
This allowed for checking the RG predictions up to the relative distances
$|1-T/T_c|\approx10^{-9}$ \cite{LipaPRB2003}, where $T_c\approx2.17$K marks the position
of the critical point ($T_c$  can be extracted out of  the specific heat data from ref.
\cite{He4RefData} with relative accuracy of about $10^{-8}$; ref.
\cite{LipaPRB2003} does not provide its value). 
These measurements, whose main outcome we briefly summarize in Fig. \ref{fig1}, 
provided the specific heat critical exponent $\alpha=-0.0127(3)$. Such a value 
 is close to the exponent that one can theoretically obtain from 
the studies of the 3D $XY$ model  \cite{VicariPRB2001}. 
It  provides significant evidence  about the
universality class of the lambda transition.
On the less optimistic side, 
these experiments are   quite  complicated in execution 
as they are performed over $200$ km above the Earth. It is thus hard to
expect that the progress in these studies can be sustained.

These results can be compared to the state of the art
experiments  in the 2D cold atom clouds placed 
in a periodic potential formed by standing laser beams (an optical lattice)
\cite{Porto2,Porto3}. These experiments aimed at determination 
of the position of the critical point  from the disappearance of the 
condensate fraction in the Mott insulator phase. They located the critical point  with 
$10$\%--$15$\% relative accuracy, which precluded experimental studies 
of the cold atom system as near the critical point as in the above-reported measurements in 
liquid $^4$He.
Moreover, the condensate fraction in these experiments 
was estimated by the relative size of the main peak in time-of-flight images. 
Such an estimation is  heuristic, as the condensate fraction is  
rather defined as the largest eigenvalue of the  single-particle density
matrix, 
which was  not 
studied in refs  \cite{Porto2,Porto3}. Finally, these references  did  not
report any measurements of the critical exponents, which excluded experimental 
identification of the universality class of the studied  system.

On the more positive side,  recent developments in the experimental manipulations of
 cold atomic samples suggest that substantial progress in the studies of phase transitions 
in these systems could be expected. For that purpose, we need efficient tools
 that can be used to extract quantitative
rather than qualitative features of phase transitions    out of experimental data.
This observation  motivates the work presented here.

The purpose of this work is to discuss a practical  scheme leading to precise determination of 
the critical points of the 3D Bose-Hubbard  model.
In addition to that, the method that we explore provides the correlation-length critical exponent,
albeit with lower accuracy.
The idea behind these studies  
was proposed in ref. \cite{BDSciRep2016}, where we tested it in the
two-dimensional 
Bose-Hubbard model. The 3D Bose-Hubbard model 
has been intensively discussed over the past decade because it provides basic description 
of the properties of cold atoms  in  optical lattices. 
Despite all this interest, precision  experimental results on its  critical
properties are hard-to-find in the literature.

The outline of this paper is the following.
We will briefly summarize below basic features of the 3D Bose-Hubbard model,
state of the art experimental and theoretical results on  positions of its  critical points, and 
our scheme  for finding them. Then, we will discuss our results
coming from Quantum Monte Carlo (QMC) simulations.

\section*{Results}

\noindent{\bf The model.}  The Hamiltonian of the 3D Bose-Hubbard model  reads \cite{FisherPRB1989,JakschPRL1998}
\ba
\label{H}
&\hat{H}=-J\sum_{\la\i, \j\ra}(\hat{a}_\i^\dag \hat{a}_\j
+ \hat{a}_\j^\dag \hat{a}_\i)
+\frac{U}{2} \sum_\i\hat{n}_\i(\hat{n}_\i-1), \\
&[\hat{a}_\i,\hat{a}_\j^\dag]=\delta_{\bf ij}, \ [\hat{a}_\i,\hat{a}_\j]=0, 
\  \hat n_\i=\hat{a}_\i^\dag\hat{a}_\i,
\ea
where $\la\i,\j\ra$ describes nearest-neighbor sites $\i$ and $\j$ in the  cubic  
lattice and periodic boundary conditions are imposed on the creation and annihilation operators. 
A physical realization of such a model, albeit with open boundary conditions, would require placing cold atoms in a
three-dimensional optical lattice enclosed in an optical box trap. The tools
needed for experimental creation of such a   trap  have been recently developed
\cite{RaizenPRA2005,HadzibabicPRL2013,Homog1,Homog2,Homog3,Homog4}.

A comprehensive  review of the properties of this model can be found in  ref. \cite{KrutitskyPhysRep2015}.
In short, its physics  depends on the filling factor $n$, i.e. the mean
number of atoms per lattice site, and the ratio of the tunneling coupling $J$
to the interaction energy $U$
\be
\eta=J/U.
\ee
We are interested in integer filling factors, for which there is a quantum
phase transition between the Mott insulator and the superfluid phase at the
critical point $\eta_c=J_c/U_c$. The
system is in the Mott
insulator  phase  for $\eta<\eta_c$ and in the superfluid phase for
$\eta>\eta_c$ \cite{FisherPRB1989}.

The critical point at unit filling factor was  theoretically studied
via perturbative expansions  \cite{MonienPRB1996,EckardtPRB2009}, 
QMC simulations \cite{SansonePRB2007},
the non-perturbative renormalization group approach \cite{DupuisPRB2011},  
and the projection operator approach \cite{SenguptaPRB2012}.
The critical points at higher filling factors  were studied 
perturbatively in ref.  \cite{MonienPRB1996} for $n=2,3$ and in
ref.  \cite{EckardtPRB2009} for an arbitrary filling factor.
The results of all above-mentioned papers can be summarized as follows
\be
\eta_c\approx\left\{
\begin{array}{ll}
0.034 & \for n=1 \\
0.02 & \for n=2 \\
0.014  & \for n=3
\end{array}
\right..
\label{Jc}
\ee

We will now summarize experimental results on the critical points of the 3D Bose-Hubbard model 
 \cite{GreinerNature2002,GerbierPRL2005,KetterlePRL2007}. 
The experiment \cite{GreinerNature2002,GreinerPhD} is done in an optical lattice of
wavelength $\lambda=852$ nm with $^{87}$Rb atoms in the $|F=2,m_F=2\ra$ state. 
Their s-wave scattering length is $5.45(26)$ nm \cite{JuliennePRL1997}. 
The position of the critical point for the
unit filling factor was estimated to correspond to lattice heights between 
$10 E_R$ and $13 E_R$, where the recoil energy $E_R$ is defined as $\hbar^2k^2/2m$ with $k=2\pi/\lambda$
and $m$ being the mass of the considered atom.  
This may be written as  
$11.5(9)E_R$,  where the standard deviation has been estimated by dividing 
 maximum uncertainty of $1.5 E_R$ by $\sqrt{3}$ \cite{sqrt3}. Using the formulas for $J$ and $U$
coefficients from   \cite{KrutitskyPhysRep2015}, we find $\eta_c$ to be  $0.04(1)$. 
The number  reported in the bracket provides one standard 
deviation due to uncertainties in the lattice height and scattering length.
It has been obtained through the uncertainty propagation formula.
Nearly identical experimental setup was used in \cite{GerbierPRL2005}. It was
found there that the  lattice heights corresponding to critical points for double and triple 
filling factors are $14.1(8)E_R$ and $16.6(9)E_R$, respectively. Applying the same
procedure as above, albeit with $\lambda=850$ nm, we get $\eta_c$ for the double and triple filling factors 
$0.021(5)$ and $0.011(2)$, respectively.
Finally, we come to ref. \cite{KetterlePRL2007}, 
where again  $^{87}$Rb atoms, but in a different hyperfine state,
are studied. It was found there that the 
superfluid--Mott insulator transition takes place 
at  $\eta_c=0.029(2)$  for the  unit
filling factor.  

To put these results in perspective, we can compare them to the mean-field
predictions, which for our system are  \cite{StoofPRA2001}
\be
\eta_c=\frac{2n+1-2\sqrt{n(n+1)}}{6}.
\ee
This yields   $\eta_c$ equal to $0.029$, $0.017$, and
$0.012$ for $n=1$, $2$, and $3$, respectively. Therefore, we  see that more 
accurate experimental results  are needed for characterization of beyond mean-field 
effects in the position of the critical points. It should be also said
that in  all above-mentioned experiments  external
harmonic trapping is imposed on the system. At the very least, it  enhances finite-size effects
making detailed
comparision between experiments and the theory based on Hamiltonian (\ref{H}) difficult. 
Such comparision is additionally complicated by the fact that 
 the  3D Bose-Hubbard model captures only the
leading-order behavior of cold atoms in  optical lattices \cite{KubaRepProg2015}.
As a result, more precise experimental results on the critical points would presumably ask for a bit more 
advanced theoretical description of the system.
Our method for locating the 
critical points should be immediately applicable to such non-standard versions of
the 3D Bose-Hubbard model.

Besides critical points, quantum phase transitions are  also characterized 
by critical exponents, which are supposed to be the same within a given 
universality class. 
The quantum 3D Bose-Hubbard model belongs to the universality class of the
classical 4D $XY$ model \cite{FisherPRB1989}. 
To the best of our knowledge, however, detailed studies of the critical properties of the
latter  model have not been presented in the literature so far. 
This is in sharp contrast to the properties of the lower-dimensional
$XY$ models, which have been studied in great detail
\cite{VicariPhysRep2002}. The difficulties presumably  arise here from  the 
complexity of  numerical studies of such  a high-dimensional model.

Furthermore, we note that  the upper critical
dimension of the $XY$ model is four. This means that the  mean-field theory,
whose  dynamical $z$ and correlation-length $\nu$ critical exponents are
given by 
\be
z=1, \ \ \nu=1/2,
\label{znu}
\ee
should  provide  the lowest-order approximation to behavior of the 4D $XY$ model. 
As it will turn out below, our relatively small system-size simulations are unable to
capture corrections to the mean-field values of the critical exponents.

\noindent{\bf The observable of interest.} We will study here variance of the  on-site atom
number distribution
\be
\Var=\la\hat n_\i^2\ra-\la\hat n_\i\ra^2,
\ee
where the site index $\i$ can be chosen arbitrarily due to the translational invariance
of our model. Such an observable can be conveniently computed with  QMC
algorithms. 
It can be also experimentally measured  \cite{TakahashiNatComm2016,3DBHexp,expNext}.
Alternatively, since  we are
actually interested in the derivative of the variance, one may focus on the  
derivative of the nearest-neighbor correlation function and use the mapping 
\be
\frac{\partial}{\partial\eta}\Var=6\eta\frac{\partial}{\partial\eta} C, \ \ 
C=\left.\la \hat a^\dag_\i\hat a_\j + a^\dag_\j\hat a_\i \ra\right|_{\la\i,\j\ra},
\ee
which  can be  obtained through the  Feynman-Hellmann theorem (it 
is strictly valid in systems that are either thermodynamically-large  or periodic).  
The nearest-neighbor correlation function  can be extracted out of the 
routinely measured  time-of-flight images in cold atom simulators 
\cite{BlochRMP2008,3DBHexp}.

\begin{figure}[t]
\includegraphics[width=\pref\textwidth,clip=true]{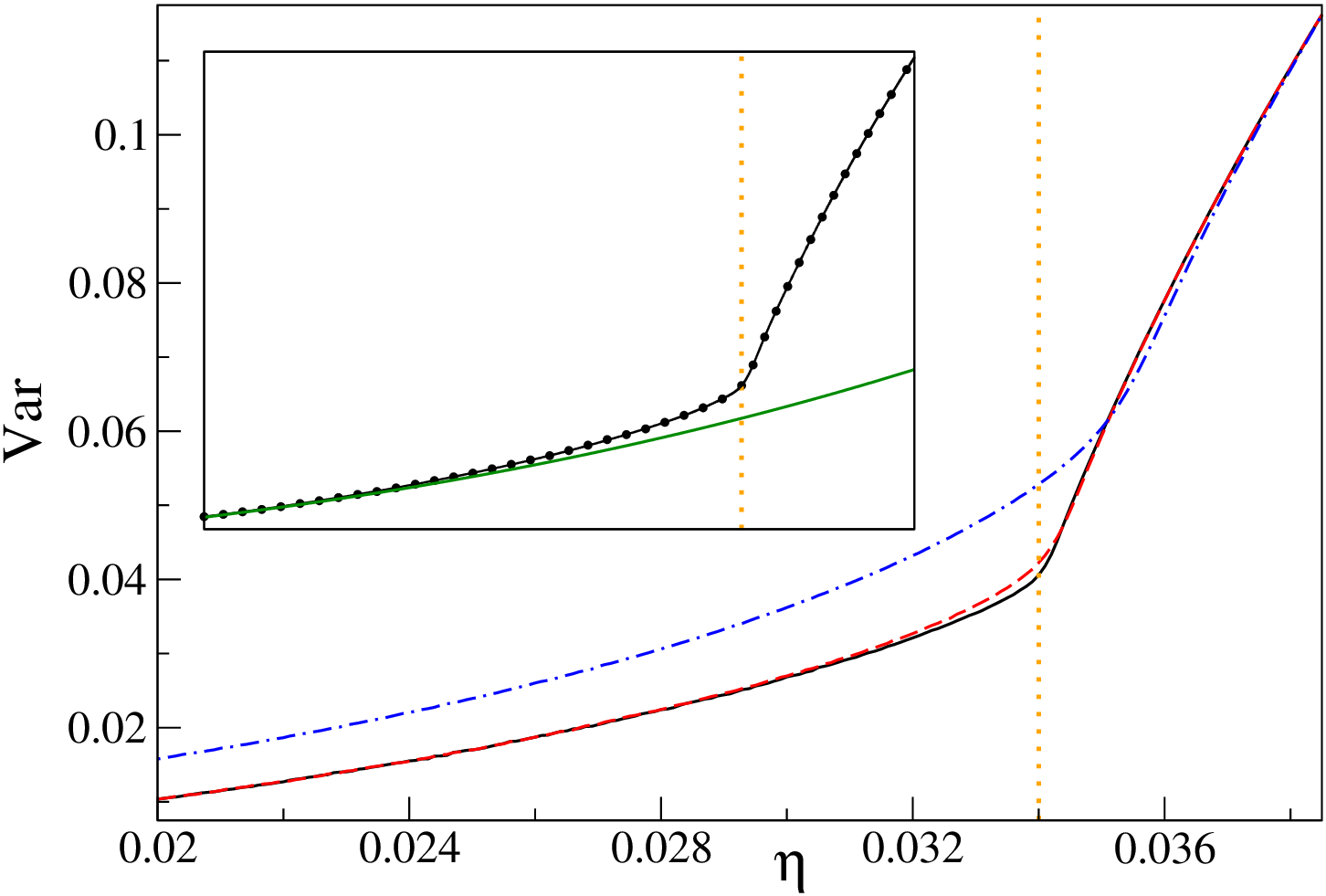}
\caption{The variance of the on-site number operator at the unit filling factor.
Main plot: solid lines show QMC data points connected by lines for 
$\kT$ equal to $0.02$ (solid black),
$0.04$ (dashed red), and $0.08$ (dashed-dotted blue).
Inset: the circles are  QMC data from the main plot for $\kT=0.02$, the solid black line is  the Pad\'{e}
approximant  fitted to the numerics, and 
the green solid line is perturbative result (\ref{var_pert}) obtained for $\kT=0$. 
The range of the
axes in the inset is the same as in the main plot.
Both here and in the following plots the vertical dotted orange lines mark the
position of the critical point (\ref{Jc}).
The  linear system size is  $L=16$. 
}
\label{fig2}
\end{figure}

The variance can
be perturbatively calculated  deeply in the Mott insulator phase at 
temperature of absolute zero
\cite{KrutitskyPhysRep2015}
\be
\la\hat n^2_\i\ra
-\la\hat n_\i\ra^2
\approx\left\{
\begin{array}{ll}
24 \eta^2+3960 \eta^4  & \for n=1 \\
72 \eta^2+33408 \eta^4 & \for n=2 \\
144 \eta^2+131400 \eta^4 & \for n=3
\end{array}
\right..
\label{var_pert}
\ee
The higher-order zero-temperature perturbative calculations of the variance in the Mott insulator phase 
of the 3D Bose-Hubbard model were  numerically performed in  comprehensive work of Teichmann {\it et
al.} \cite{EckardtPRB2009}.

\noindent{\bf Quantum Monte Carlo simulations.}
We perform  QMC simulations, which we briefly describe in the Methods section
(see \cite{PolletRepProgPhys2012} for a cold-atom-oriented  review of this subject). 
This   allows us to  study physics on the
superfluid side of the transition, where dependence of the  variance on $\eta$
is most interesting for our purposes. Additionally, this approach allows us to get
nonzero-temperature results, which is of interest from the experimental perspective.

\begin{figure}[t]
\includegraphics[width=\pref\textwidth,clip=true]{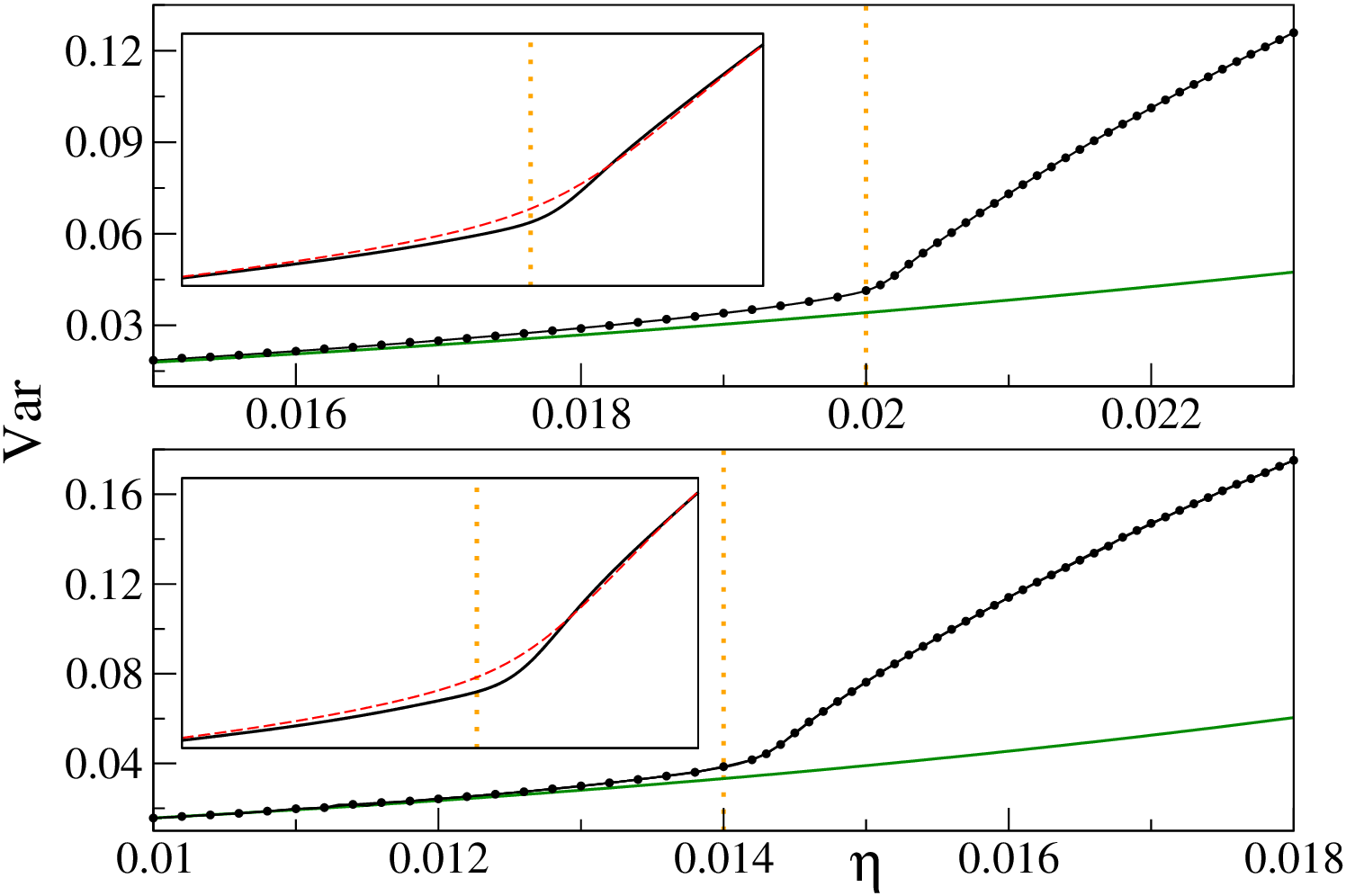}
\caption{The variance of the on-site atom number operator at  filling factors
$n=2$ (upper panel) and $3$ (lower panel).
Upper panel: the solid black line shows the Pad\'{e} approximant for the linear
system size  $L=16$. 
The inset illustrates 
differences between Pad\'{e} approximants for $L=16$ (black line)
and $L=8$ (red dashed line)  near the critical point. 
The range of the
horizontal (vertical) axis in the inset is $0.0185\le\eta\le0.021$
($0.03\le\Var\le0.075$). 
Lower panel: the solid black line depicts the Pad\'{e} approximant for $L=12$. 
The inset shows differences around  the critical point
between  Pad\'{e} approximants for $L=12$ (black line)
and $L=6$ (red dashed line).
The range of the
horizontal (vertical) axis in the inset is $0.012\le\eta\le0.0155$
($0.022\le\Var\le0.1$). 
The solid green line in both panels shows perturbative results
(\ref{var_pert}) for $\kT=0$. All other results, including  circles showing
QMC data, are obtained for $\kT=0.02$.
}
\label{fig3}
\end{figure}

We perform our studies  in lattices of size 
\be
L^3,
\label{L3}
\ee
where the linear system size $4\le L\le16$ for $n=1,2$ and $4\le L\le12$ for $n=3$.  
Most of the time, we investigate systems at  temperature 
$\kT=0.02$, where $k_B$ is the Boltzmann constant. Such  temperature can be converted to Kelvins by
considering  typical experimental conditions. To do so,
one may assume  that the lattice of wavelength  $\lambda=532$ nm is
populated by either $^{87}$Rb or $^{174}$Yb atoms having the s-wave scattering
lengths of $5.45$ nm and $5.56$ nm, respectively \cite{JuliennePRL1997,YbaPRA2017}. 
Using then  formulas from \cite{KrutitskyPhysRep2015}, we 
find that $U_c/E_R$   for $n=1,2,3$ equals 
$0.47$, $0.54$,  $0.59$ for $^{87}$Rb  and
$0.48$, $0.55$, $0.60$  for
$^{174}$Yb. These numbers have been obtained by assuming that critical
points are given by (\ref{Jc}). 
Since we work near critical points, we may take $U_c/k_B$ as the unit of temperature. In nano
Kelvins, it equals 
$184$, $211$, $230$ 
($93$, $107$, $117$)
for $n=1,2,3$ in  $^{87}$Rb ($^{174}$Yb). 
We have checked that the same results can be obtained by generating  Wannier functions and then
integrating them over in order to get  $J$ and $U$ coefficients
\cite{JakschPRL1998}.

We see from these calculations that  temperature $\kT=0.02$ corresponds to
a few nano Kelvins in  typical  rubidium and ytterbium systems.  
While such  low temperatures are   certainly experimentally challenging, 
it does not mean that our studies are completely free from nonzero-temperature 
corrections. For example, the 3D Bose-Hubbard model was studied 
through QMC simulations in ref. \cite{SansonePRB2007} at temperature about 
twenty  times  smaller than our $\kT=0.02$. These studies were done at the unit
filling factor in systems, whose sizes were similar   to the ones used by us. The critical
point was extracted from  finite-size scaling of the excitation gap.
The relative difference between the position of the critical point
found in our work and the one from ref. \cite{SansonePRB2007} is about $0.5\%$. 
We will thus skip systematic discussion of nonzero-temperature effects in our
computations. 
It should be stressed, however, that our approach to finding 
critical points can be applied to ``warmer'' systems as well, where 
nonzero-temperature scaling  analysis can be deployed \cite{ContinentinoBook}.

\begin{figure}[t]
\includegraphics[width=\pref\textwidth,clip=true]{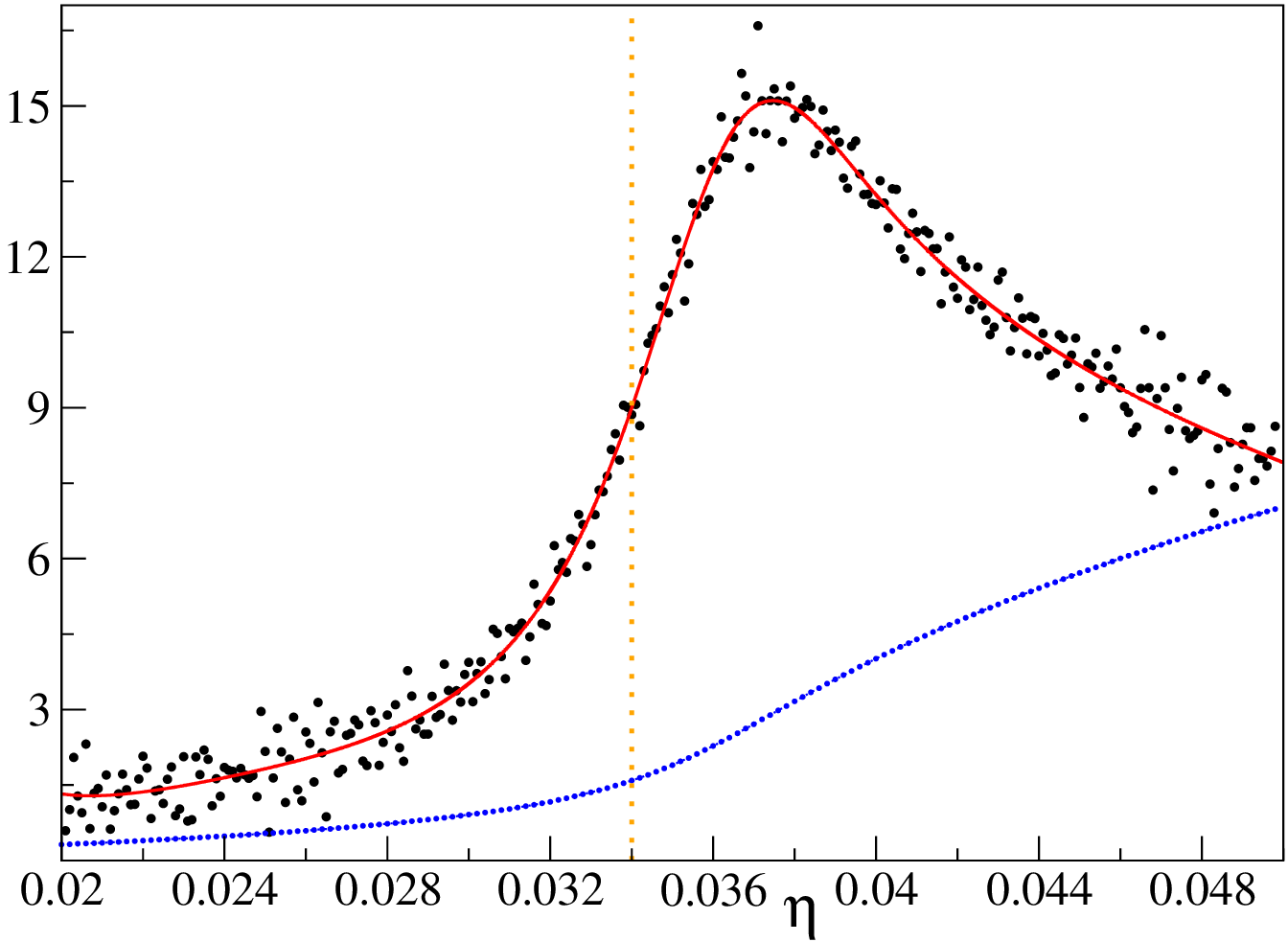}
\caption{The first derivative of the variance for the unit filling factor: raw data vs.
the Pad\'{e} approximant result. Black dots: the central difference numerical
derivative of the QMC data for the variance. 
Three hundred equally-spaced 
data points were used to compute the derivative. The value represented by every other 
such data point is multiplied by a factor of thirty and depicted as a small blue dot in the plot.
The solid red line is the derivative of the Pad\'{e} approximant  that has been fitted to the QMC numerics. 
Computations  for this plot have been  done for $\kT=0.02$ and the linear system size
$L=4$.
}
\label{fig4}
\end{figure}

The variance for the filling factors $n=1$ and $n=2,3$ is
presented in Figs \ref{fig2} and \ref{fig3}, respectively. 
We see there
its steep increase around the expected positions of  critical points (\ref{Jc}).
To locate the points, where  changes of the variance proceed most rapidly,
we compute the first derivative of the variance. Such a  procedure encounters a
technical problem: the derivative is sensitive to fluctuations of
 data that is being differentiated (Fig. \ref{fig4}).
Therefore,  such data has to be smoothed first, 
which we do by fitting the Pad\'{e} approximant
of order $(m,s)$ \cite{PadeBook} 
\be
\Var(\eta)=\frac{\sum_{i=0}^m A_i \eta^i}{1+\sum_{i=1}^s B_i\eta^i}.
\label{Pade}
\ee
Usefulness of this procedure is illustrated in Fig. \ref{fig4}, where 
 we see that the derivative of the Pad\'{e} approximant
provides a smooth curve that can be easily subjected to detailed analysis. 
We mention in passing that the very same procedure could be applied to data for the variance coming from 
experimental measurements.

Fitting  noisy data with  (\ref{Pade}) requires the   approximation order   to
be adapted to the problem. 
While small orders   may result in a bad fit due to 
insufficient number of fitting parameters, 
choosing too large orders causes problems as well. In the  latter case,
such extra flexibility  leads to  reproduction of 
noise-induced fluctuations of  data points  
 instead of averaging the fluctuations  out.

\begin{figure}[t]
\includegraphics[width=\pref\textwidth,clip=true]{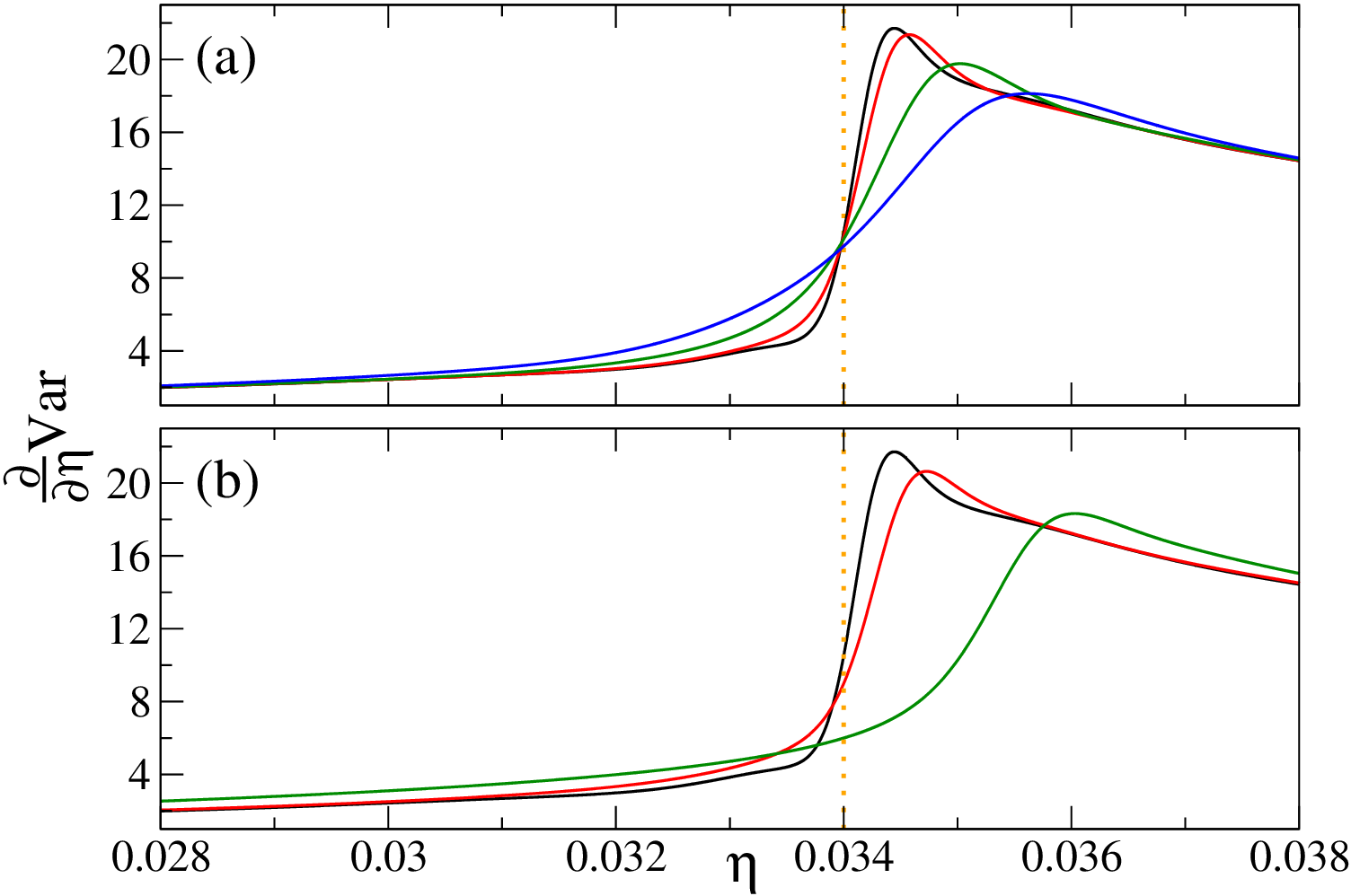}
\caption{The first derivative of the  variance of the on-site atom 
number operator at the unit filling factor. The derivative is 
computed from Pad\'{e} approximants to QMC data.
Panel (a): system-size dependence for $\kT=0.02$. 
The curves from top to bottom correspond to linear system sizes $L$ equal to $16$ (black), $12$ (red), 
$8$ (green),  and $6$ (blue), respectively.
Panel (b): temperature dependence for  $L=16$. 
The curves from top to bottom correspond to $\kT$ equal to $0.02$ (black), $0.04$ (red), 
and $0.08$ (green), respectively.
}
\label{fig5}
\end{figure}

\begin{figure}[t]
\includegraphics[width=\pref\textwidth,clip=true]{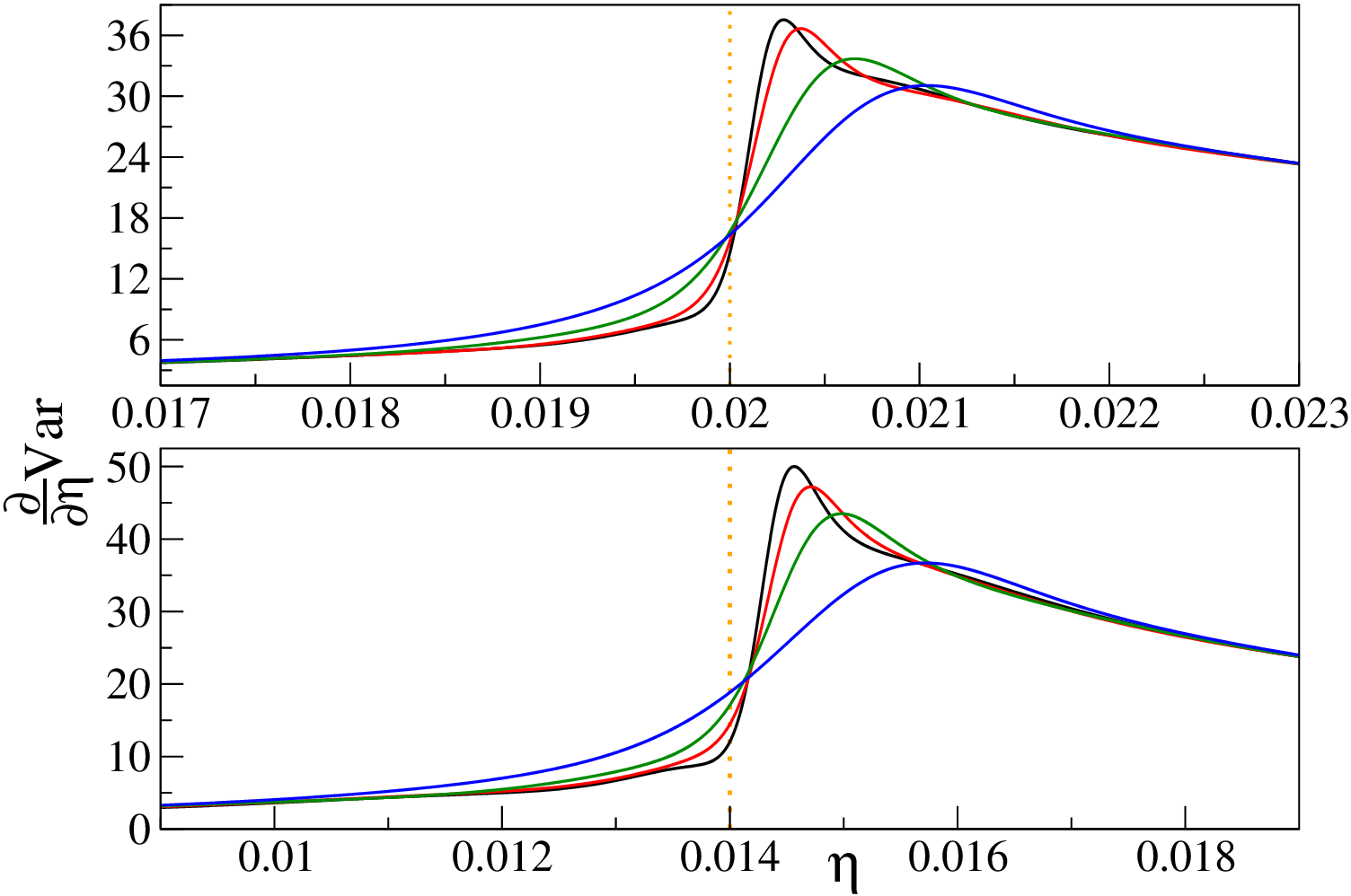}
\caption{The first derivative of the variance of the on-site atom number operator at  filling factors
$n=2$ (upper panel) and $3$ (lower panel).
Upper panel: the curves from top to bottom are  for linear system sizes $L$ equal to $16$ 
(black), $12$ (red), $8$ (green), and $6$ (blue).
Lower panel: the curves from top to bottom  are prepared for $L$ equal to $12$ (black), $8$ (red), $6$ (green),
and $4$ (blue).
Derivatives of  Pad\'{e} approximants  are displayed in both panels, all
calculations are done for $\kT=0.02$.\\
}
\label{fig6}
\end{figure}

\begin{figure}[t]
\includegraphics[width=\pref\textwidth,clip=true]{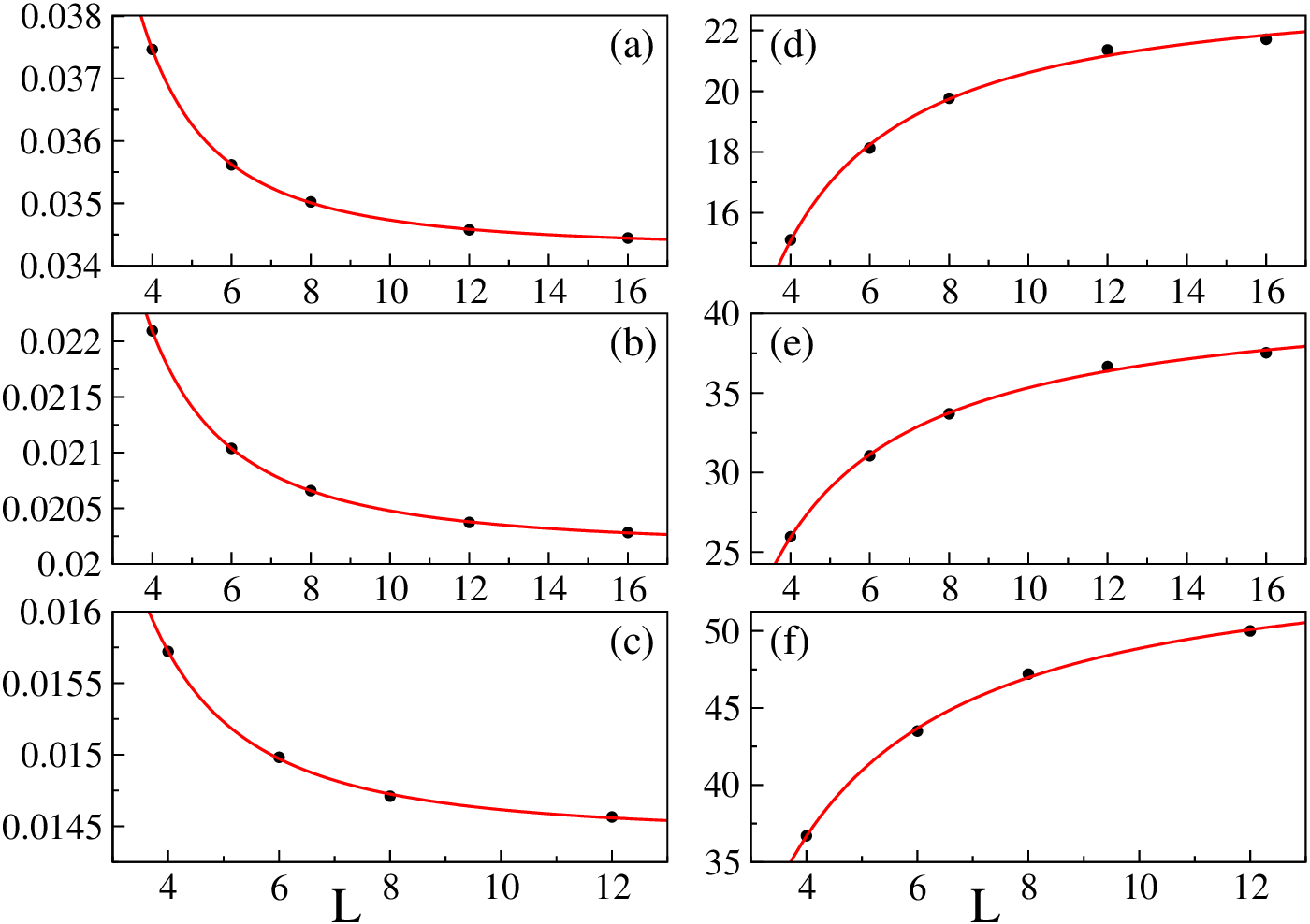}
\caption{Properties of the maximum of the first derivative of the variance.
The black  dots are extracted from  Pad\'{e} approximants to QMC data, the red solid lines
are nonlinear fits (\ref{nlinfit}) and (\ref{nlinfit1}). 
Panels (a), (b), and (c) show $\eta^*$  for filling factors $n=1$, $2$, and $3$, respectively.
The fitted curves are:
$0.03427(2)+0.060(3) L^{-2.12(4)}$, 
$0.020144(8)+0.0282(8) L^{-1.93(2)}$, and 
$0.01444(4)+0.026(6) L^{-2.2(2)}$ (top to bottom).
Panels (d), (e), and (f) show $\dVar(\eta^*)$ for filling factors $n=1$, $2$, and $3$, respectively. 
The fitted curves are:
$23.6(7)-42(8) L^{-1.2(2)}$, 
$42(1)-62(8) L^{-1.0(1)}$,
and $55(2)-94(18) L^{-1.2(2)}$ (top to bottom).
We report here one standard deviation in the brackets next to the fitted coefficients.
The fitting is done with  NonlinearModelFit function from \cite{Mathematica}. 
All results are for $\kT=0.02$.
}
\label{fig7}
\end{figure}

\begin{figure}[t]
\includegraphics[width=\pref\textwidth,clip=true]{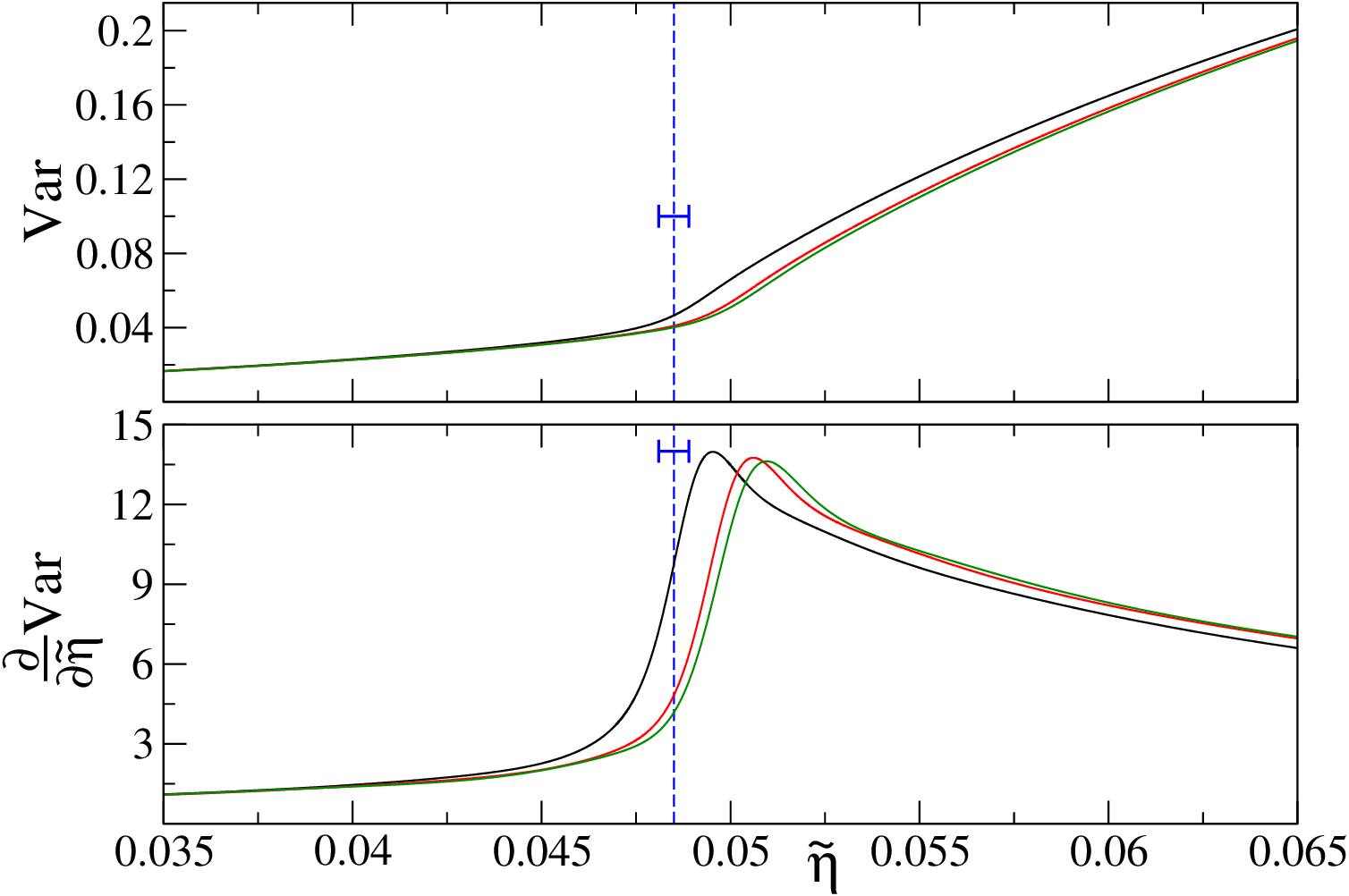}
\caption{The variance of the on-site atom number operator  and its derivative
as a  function of  rescaled variable (\ref{teta}).
The solid black, red, and green lines--top to bottom in the upper panel and
left to right in the lower panel--present  Pad\'{e}
approximants  to our QMC numerics   for filling 
factors $n=1$, $2$, and $3$, respectively. The critical values of
$\tilde\eta$, which can be obtained from (\ref{Jc}),
are approximately $0.0481$, $0.0490$ and $0.0485$ for filling factors $n=1$,
$2$ and $3$, respectively. The vertical dashed blue lines are placed at
$\tilde\eta$ equal to $0.0485$, the mean of the above-reported numbers.
The spread of the positions of the critical points for
different filling factors around such mean,
$\pm0.0004$, is marked by the horizontal ``error bars''.
All results are for the linear system size $L=8$ and temperature $\kT=0.02$.
}
\label{fig8}
\end{figure}

Choosing the optimal order  of the Pad\'{e} approximant is not difficult in our computations. 
Indeed, we have found that for every combination of $(L,n,T)$ parameters,
there exists a stability island 
 in the set of all reasonable approximation orders.  
By taking  the order of approximation  within the island, 
 stable results for the variance and its derivative are obtained. 
We have found that typically our QMC data sets can be reasonably fitted with  $8\le m,s \le9$.
We have also found that by considering  a denser numerical grid, or by reducing the 
QMC noise by increasing the sample size, stable results can be obtained. 
We have applied this strategy for creation of Figs \ref{fig2}--\ref{fig8}, where
Pad\'{e} approximants of the  fixed $(8,8)$ order are employed.

The system-size and temperature dependence of $\dVar$ is presented in
Figs  \ref{fig5} and \ref{fig6} for the filling factors $n=1$ and $n=2,3$,
respectively. The first thing we notice there is the lambda-shape of the plots
reminiscent of the specific-heat plot in liquid $^4$He (Fig. \ref{fig1}).

Then, we find   that  the first derivative of the
variance has a maximum near the critical point on the superfluid side of the transition,
say at $\eta^*$.
We see that $\eta^*$ shifts towards the critical point as the system
size increases. The same is observed when temperature decreases. Moreover,
$\dVar(\eta^*)$ grows with the system size and inverse
temperature.
All these observations suggest that the maximum of the derivative of the
variance encodes the
position of the critical point. This is not the first time when the derivative
of an experimentally-accessible quantity is  used for finding the critical
point of the 3D Bose-Hubbard model. 
Indeed, the derivative of  experimentally-measured visibility of the 
time-of-flight interference pattern
was used for such a purpose as well \cite{GerbierPRL2005}.

More quantitatively, we study the position of the maximum of $\dVar$
by fitting 
\be
a+b L^{-c}
\label{nlinfit}
\ee
to  numerical results. The idea here is that the parameter $a$ estimates the
position of the maximum in the thermodynamically-large system ($c>0$).
The typical fits that we perform are shown in Figs
\ref{fig7}a--c, where the positions of the maxima
 have been  extracted  from  Pad\'{e} approximants of order $(8,8)$. 
To check sensitivity of these results to the order of  approximants
 (\ref{Pade}), we  have done  calculations for 
\be
7\le m,s\le9.
\label{ms}
\ee
We have obtained 
\be
a=\left\{
\begin{array}{ll}
0.03430\pm0.00008  & \for n=1 \\
0.02020\pm0.00006  & \for n=2 \\  
0.01447\pm0.00003 & \for n=3
\end{array}
\right.,
\label{a}
\ee
where the error bars are chosen to capture all the results in the parameter
range given by  (\ref{ms}).
Standard deviations of fitted coefficients for   $m=s=8$ are 
typically a bit smaller (Fig. \ref{fig7}). 
A quick look at (\ref{Jc}) reveals that these results
provide  positions of  critical points, which makes us confident that $\eta^*$
converges to $\eta_c$ in the thermodynamic limit.

For the $c$ coefficient under constraint  (\ref{ms}), we have got 
\be
c=\left\{
\begin{array}{ll}
2.11\pm0.08 & \for n=1 \\ 
1.97\pm0.08 & \for n=2 \\ 
2.23\pm0.05 & \for n=3
\end{array}
\right.,
\label{c}
\ee
which can be explained through the standard finite-size scaling argument \cite{Cardy1,Cardy2}.
Namely, we assume that near the critical point the system properties depend on the linear system size
and the ratio between the linear system size and the correlation length $\xi$, say 
\be
\dVar=h(L) f\B{\frac{L}{\xi}}, \ \ \xi\sim|\eta-\eta_c|^{-\nu},
\label{fss}
\ee
where $f$ is a non-singular scaling function. According to this ansatz, 
the position of the extremum of $\dVar$  scales
\be
|\eta^*-\eta_c|\sim L^{-1/\nu}
\label{finiteLnu}
\ee
and we see that the function $h(L)$ captures system-size dependence of $\dVar$
at $\eta^*$.
The exponent in this equation matches the $c$ parameter if we use the
mean-field value of the critical exponent $\nu$  (\ref{znu}). More accurate studies
are needed for checking if there are beyond-mean-field corrections to the critical
exponent $\nu$ in the 3D Bose-Hubbard model. We believe that the key limitations here come 
from the relatively small system sizes that can be numerically handled.

Finally, for the sake of completeness, we provide the results for the
fitting parameter $b$ again under the variation of the order of Pad\'{e}
approximation (\ref{ms})
\be
b=\left\{
\begin{array}{ll}
0.059\pm0.005 & \for n=1 \\ 
0.029\pm0.002  & \for n=2 \\ 
0.028\pm0.002  & \for n=3
\end{array}
\right..
\label{b}
\ee

Next, we will discuss scaling of $\dVar(\eta^*)$ with the linear system size. We fit
\be
A+BL^{-C}
\label{nlinfit1}
\ee
to  numerics. Typical results that we obtain are presented in Figs
\ref{fig7}d--f, where again  Pad\'{e} approximants of order $(8,8)$ have been
employed. Next, we quantify influence  of the order of Pad\'{e}
approximation on these results. Proceeding similarly as with 
(\ref{a}), (\ref{c}), and (\ref{b}), we get the results summarized in Table \ref{tab}.
All of them  suggest that $\dVar(\eta^*)$ slowly increases with the linear system
size reaching a finite value in the thermodynamic limit. This has an
interesting  consequence  that can be readily spotted   in Figs \ref{fig5} and \ref{fig6}. 

Namely, we see
that  the curves showing $\dVar$ for different  
system sizes at constant temperature cross near the critical point. This can be explained by the finite-size
scaling ansatz (\ref{fss}), if we note that  $\dVar(\eta_c)\sim h(L)$ and 
take into account that 
$h(L)$ weakly depends on the linear system size reaching 
a finite value in the thermodynamic limit. The latter remark
follows from the fact that $h(L)$ is proportional to $\dVar(\eta^*)$, which we
have just discussed. We mention in passing that similar-looking crossing of 
curves  near the critical point was used  for finding the position of the
critical point from QMC data for the excitation gap \cite{SansonePRB2007}.

\begin{table}[t]
\begin{tabular}{ l | c | c | c}
   & A & B & C \\
  \hline
  $n=1$ & $23.3\pm0.6$   &   $-44\pm4$       & $1.2\pm0.1$	 \\
  \hline  
  $n=2$ & $41\pm1$   &   $-64\pm4$   & $1.1\pm0.1$	 \\
  \hline
  $n=3$ & $54\pm1$   &   $-101\pm7$     & $1.3\pm0.1$	  \\
\end{tabular}
\caption{Coefficients obtained by  fitting (\ref{nlinfit1}) to
$\dVar(\eta^*)$. Error bars capture spread of the results due to the varying  order of Pad\'{e}
approximation (\ref{ms}). QMC results  for $\kT=0.02$ have been used to prepare this
table, the same system sizes as in Fig. \ref{fig7} have been employed in the
fitting.}
\label{tab}
\end{table}

Further insights into $\dVar$  can be obtained by setting $\kT=0$ and using 
the Feynman-Hellmann theorem  to arrive at  \cite{BDSciRep2016}
\be
\frac{\partial}{\partial\eta}\Var=-2\eta\frac{\partial^2}{\partial\eta^2}\B{\frac{\cal E}{U}},
\label{aaa}
\ee
where $\cal E$ is the ground-state energy per lattice site. This
expression is closely linked to the one  for  specific heat
oftentimes  studied in the context of classical phase transitions (see e.g. 
Fig. \ref{fig1} and the discussion around it). Indeed, 
specific heat per lattice site can be 
written as \cite{Baxter}
\be
-T\frac{\partial^2}{\partial T^2}\cal F,
\label{bbb}
\ee
where $\cal F$ is the free energy per lattice site. Its singular part is
typically assumed to scale as $|T-T_c|^{2-\alpha}$, where $\alpha$ is the
specific heat critical exponent. A quick look at (\ref{aaa}) and (\ref{bbb})
reveals the mapping between the two expressions.  It is then unsurprising 
that the singular part of $\cal E$ is usually  assumed  to scale as
\be
|\eta-\eta_c|^{2-\alpha}.
\label{2malpha}
\ee
The exponent $\alpha$ is linked to the $z$ and  $\nu$ critical
exponents through  the quantum hyperscaling relation 
\be
\alpha = 2- \nu(d + z),
\label{hyper}
\ee
where  $d$ is  the system's  dimensionality \cite{ContinentinoBook}.

Combining (\ref{aaa}) with (\ref{2malpha}),  one gets
$\dVar\sim|\eta-\eta_c|^{-\alpha}$
in the infinite system, which would imply that $h(L)\sim L^{\alpha/\nu}$. As a
result, $\alpha=0$ would be compatible with our numerics in the large-$L$
limit. Such a value can be
obtained by putting mean-field critical exponents (\ref{znu}) into (\ref{hyper}).
There are, however, at least two  reasons to be cautious here.

First, the upper critical dimension of the  Bose-Hubbard model is
three and so it is expected that there will be corrections 
to the mean-field scaling laws. As a result,
it is unclear to us what is the actual value of $\alpha$. 

Second, even if $\alpha$ would be zero, 
the  presence of logarithmic singularities in the derivatives of the ground-state 
energy could not be ruled out without detailed analysis. 
For example, such a situation  takes place in the one-dimensional 
quantum Ising model, where $\alpha=0$ due to $z=\nu=d=1$
\cite{Lieb1961,PfeutyAnn1970}. 
The singular part of its ground-state energy per lattice site $\EIsing$ turns out to be proportional to 
$(g-g_c)^2\ln|g-g_c|$ in the thermodynamically-large chain, 
where $g$ is the magnetic field driving the transition
and $g_c$ is the critical point. As a result, $\dwaE$  diverges logarithmically with $|g-g_c|$.
 In the finite  chain, there is the  extremum  of 
$\dwaE$ at say $g^*$. When the system size increases, $g^*\to g_c$ and 
$|\dwaE(g^*)|\to\infty$. 
The latter property differs from  what we seem to observe in the 3D Bose-Hubbard
model.

One possible explanation of our  puzzling observation that   $\dVar(\eta^*)$ 
reaches a finite value in the thermodynamic limit
might be that the system sizes that we consider are much
too limited rendering our extrapolations  unreliable. Still, the use of our
fitting results for interpolation purposes should be very well justified and useful.

Finally, we would like to briefly discuss dependence  of our results on the
filling factor $n$. The idea that we explore  here comes from Teichmann {\it
et al.} 
\cite{EckardtPRB2009}, where it was found through perturbative studies 
that deeply in the Mott insulator phase the variance of the on-site atom number
operator is a function of
\be
\tilde{\eta}=\sqrt{n(n+1)}\eta.
\label{teta}
\ee
This implies that $\dVartilde$ should be a function of (\ref{teta}) as well.

Our QMC numerics, which we present in Fig. \ref{fig8},  perfectly follows  these predictions away from 
the critical point on 
the Mott insulator side of the transition. 
We also see in this figure that  the mapping $\eta\mapsto\tilde\eta$ fails a bit
in the superfluid phase for low-$n$ data that we explore. 
Nonetheless, judging from quite good 
overlap between the $n=2$ and $3$ results, 
it is reasonable to expect that 
the  mapping  will
be accurately supported by numerical simulations  in both phases in the limit of $n\gg1$. 
Further studies are needed for establishing  this observation.

\section*{Discussion}
We have  studied equilibrium properties of the 3D  Bose-Hubbard
model focusing our attention on the variance of the on-site atom number operator
and its derivative with respect to the parameter driving the superfluid-Mott
insulator transition. Our results have been obtained 
in systems with the mean number of atoms per lattice site equal to one, two, and
three. They come from   Quantum Monte Carlo simulations.

The key finding of this work is that the  derivative of the variance has a
pronounced maximum   close to  critical points. 
For example, in a very small lattice of linear size
$4$, when the number of atoms equals the number of lattice sites,
the position of the maximum estimates the position of the 
critical point with $10\%$ relative accuracy (Fig. \ref{fig4}).
The mismatch between the two decreases quadratically with the linear system  size 
and it can be further suppressed  by simple  extrapolation to the
thermodynamic limit.

Besides discussing the position of the critical point, which is an interesting albeit
non-universal feature, we have found that even in small systems   the critical
exponent $\nu$ can be extracted from the finite-size shift of the maximum
of the derivative of the variance. This is interesting because 
knowledge of this  exponent can provide important information
about the universality class of the 4D $XY$ model.

This is the least-studied universality class of the $XY$ model. Limited
knowledge of its properties  stems from numerical shortcomings, clearly seen in our
work, and difficulties in finding  physical systems, where it can be experimentally approached.
The latter can be done in condensed matter and atomic physics setups.
In the condensed matter context,  it was proposed that some properties of either    
strongly underdoped cuprate superconductors
or $^4$He in nanoporous media can be  captured by  the 4D $XY$  universality class 
\cite{FranzPRL2006,BrounPRL2007,EggelPRB2011}. 
In the atomic physics context, cold atoms in a three-dimensional optical lattice 
are the best example of a system whose scaling properties should mirror those of the
4D $XY$ model. 

We view  cold atom setups, simulating the
3D Bose-Hubbard model,
as the cleanest and  most promising platform for future  quantitative studies of the
4D $XY$ universality class. In fact, measurements of the on-site atom number
fluctuations in the 3D Bose-Hubbard systems have been recently reported
\cite{3DBHexp,expNext}.
Direct comparision of these results to our findings is difficult because 
setups studied  in \cite{3DBHexp,expNext} 
are non-uniform due to the external trapping potential 
adding a local chemical-potential-like term to  Hamiltonian (\ref{H}). We are
hopeful, however, that blending  of  techniques presented in these
references  with the recent optical
box trapping advances  can lead to successful creation of the homogeneous 3D Bose-Hubbard
quantum simulator.
Such a system could  be large-enough to overcome small-size limitations 
plaguing  numerical simulations. As a matter of fact, quantum simulators,
at the very least, are supposed to do just that.

\noindent{\bf Methods}

\noindent We use the Directed Worm Algorithm from  the  ALPS software package 
\cite{alps1,alps2}.
This algorithm  samples the path-integral representation of a density matrix of a  
grand  canonical ensemble (GCE) with configurations called worldlines.  
Since we work with  systems having  fixed filling factor $n$, 
computation of any average in a lattice of the linear size $L$  requires rejecting those worldlines,
where the total numer of particles differs from $n L^3$.
To improve the  sample count of the remaining fraction, the chemical potential is 
adjusted to set   the expected GCE density of the system  to $n$ particles per
site. 
The statistical error of the determined variance is significantly reduced by
adopting periodic boundary conditions,  where observables do not depend on the lattice site.
As a result, the variance of the on-site atom number operator
can be averaged  over the resulting ensemble and over all $L^3$ lattice sites, which is
exactly what we do.

Due to the amount of computational power needed, our 
QMC simulations are limited to system sizes and temperatures discussed
below equation (\ref{L3}). Early symptoms of these limitations can be spotted 
in  Figs \ref{fig5} and \ref{fig6}. We see there that for the largest systems considered,
 there is a small warp in the derivative of the variance 
slightly to the left of the dotted lines marking positions of  critical points.
Therefore, it is  important to check   that positions of the maxima, which
we extensively study, are stable under  change of  parameters of our QMC simulations. 
Several tests are thus performed. First, 
we vary the total number of later averaged worldlines reaching typically the level of $10^7$ to $10^8$.
Second, when generating the  worldlines, 
only every $m$-th trajectory is included into the final ensemble 
(if it additionally contains $nL^3$ particles), to ensure that subsequent 
worldlines in the ensemble are independent from each other. 
We check sensitivity of our results to this so-called skip $m$ parameter by changing it from $64$ to
$256$. Third,  we  fit  Pad\'{e} approximants to 
data sets that differ from each other by range and grid density of the  parameter $\eta$.
All these tests make us confident that the results that we present for the maxima
of the derivative of the variance are  well-converged.

\noindent{\bf Acknowledgments}

\noindent 
We thank Yoshiro Takahashi for a conversation about ref. \cite{3DBHexp}.
We also 
thank Tama Ma for discussions  about the details of the ALPS implementation of the Directed Worm Algorithm.
OAP and BD were supported by the Polish National Science Centre (NCN) grant DEC-2016/23/B/ST3/01152.
MŁ was supported by  the Polish National Science Centre (NCN) project 2016/23/D/ST2/00721. 
Numerical computations in this work were supported in part by PL-Grid Infrastructure.


\begin{thebibliography}{52}
\expandafter\ifx\csname natexlab\endcsname\relax\def\natexlab#1{#1}\fi
\expandafter\ifx\csname bibnamefont\endcsname\relax
  \def\bibnamefont#1{#1}\fi
\expandafter\ifx\csname bibfnamefont\endcsname\relax
  \def\bibfnamefont#1{#1}\fi
\expandafter\ifx\csname citenamefont\endcsname\relax
  \def\citenamefont#1{#1}\fi
\expandafter\ifx\csname url\endcsname\relax
  \def\url#1{\texttt{#1}}\fi
\expandafter\ifx\csname urlprefix\endcsname\relax\def\urlprefix{URL }\fi
\providecommand{\bibinfo}[2]{#2}
\providecommand{\eprint}[2][]{\url{#2}}

\bibitem[{Sac({\natexlab{a}})}]{Sachdev}
\bibinfo{note}{Sachdev, S. {\it {Q}uantum {P}hase {T}ransitions} (Cambridge
  University Press, 2011).}

\bibitem[{Sac({\natexlab{b}})}]{SachdevToday}
\bibinfo{note}{Sachdev, S. \& Keimer, B. Quantum criticality. {\it Phys. Today}
  {\bf 64}, 29 (2011).}

\bibitem[{Con()}]{ContinentinoBook}
\bibinfo{note}{Continentino, M. {\it Quantum Scaling in Many-Body Systems: An
  Approach to Quantum Phase Transitions} (Cambridge University Press, 2nd
  edition, 2017).}

\bibitem[{Fey()}]{Feynman}
\bibinfo{note}{Feynman, R. P. Simulating physics with computers. {\it Int. J.
  Theor. Phys.} {\bf 21}, 467 (1982).}

\bibitem[{Gre({\natexlab{a}})}]{GreinerNature2002}
\bibinfo{note}{Greiner, M., Mandel, O., Esslinger, T., H\"ansch, T.W. \& Bloch,
  I. Quantum phase transition from a superfluid to a Mott insulator in a gas of
  ultracold atoms. {\it Nature} {\bf 415}, 39 (2002).}

\bibitem[{Lip()}]{LipaPRB2003}
\bibinfo{note}{Lipa, J. A., Nissen, J. A., Stricker, D. A., Swanson, D. R. \&
  Chui, T. C. P. Specific heat of liquid helium in zero gravity very near the
  lambda point. {\it Phys. Rev. B} {\bf 68}, 174518 (2003).}

\bibitem[{Lew()}]{LewensteinAdv}
\bibinfo{note}{Lewenstein, M., Sanpera, A., Ahufinger, V., Damski, B., Sen(De),
  A. \& Sen, A. Ultracold atomic gases in optical lattices: mimicking condensed
  matter physics and beyond. {\it Adv. Phys.} {\bf 56}, 243 (2007).}

\bibitem[{Blo()}]{BlochRMP2008}
\bibinfo{note}{Bloch, I., Dalibard, J. \& Zwerger, W. Many-body physics with
  ultracold gases. {\it Rev. Mod. Phys.} {\bf 80}, 885 (2008).}

\bibitem[{Car({\natexlab{a}})}]{Cardy2}
\bibinfo{note}{Cardy, J. {\it Scaling and Renormalization in Statistical
  Physics} (Cambridge University Press, Cambridge, 2002).}

\bibitem[{He4()}]{He4RefData}
\bibinfo{note}{Donnelly, R. J. \& Barenghi, C. F. The observed properties of
  liquid helium at the saturated vapor pressure. {\it J. Phys. Chem. Ref. Data}
  {\bf 27}, 1217 (1998).}

\bibitem[{Vic({\natexlab{a}})}]{VicariPRB2001}
\bibinfo{note}{Campostrini, M., Hasenbusch, M., Pelissetto, A., Rossi, P. \&
  Vicari, E. Critical behavior of the three-dimensional $\mathrm{XY}$
  universality class. {\it Phys. Rev. B} {\bf 63}, 214503 (2001).}

\bibitem[{Por({\natexlab{a}})}]{Porto2}
\bibinfo{note}{Spielman, I. B., Phillips, W. D. \& Porto, J. V. Condensate
  fraction in a 2D Bose gas measured across the Mott-insulator transition. {\it
  Phys. Rev. Lett.} {\bf 100}, 120402 (2008).}

\bibitem[{Por({\natexlab{b}})}]{Porto3}
\bibinfo{note}{Jim\'enez-Garc\'{\i}a, K. {\it et al.} Phases of a
  two-dimensional Bose gas in an optical lattice. {\it Phys. Rev. Lett.} {\bf
  105}, 110401 (2010).}

\bibitem[{BDS()}]{BDSciRep2016}
\bibinfo{note}{Łącki, M., Damski, B. \& Zakrzewski, J. Locating the quantum
  critical point of the Bose-Hubbard model through singularities of simple
  observables. {\it Sci. Rep.} {\bf 6}, 38340 (2016).}

\bibitem[{Fis()}]{FisherPRB1989}
\bibinfo{note}{Fisher, M. P. A., Weichman, P. B., Grinstein, G. \& Fisher, D.
  S. Boson localization and the superfluid-insulator transition. {\it Phys.
  Rev. B} {\bf 40}, 546 (1989).}

\bibitem[{Jak()}]{JakschPRL1998}
\bibinfo{note}{Jaksch, D., Bruder, C., Cirac, J. I., Gardiner, C. W. \& Zoller,
  P. Cold bosonic atoms in optical lattices. {\it Phys. Rev. Lett.} {\bf 81},
  3108 (1998).}

\bibitem[{Rai()}]{RaizenPRA2005}
\bibinfo{note}{Meyrath, T. P., Schreck, F., Hanssen, J. L., Chuu, C.-S. \&
  Raizen, M. G. Bose-Einstein condensate in a box. {\it Phys. Rev. A} {\bf 71},
  041604 (2005).}

\bibitem[{Had()}]{HadzibabicPRL2013}
\bibinfo{note}{Gaunt, A. L., Schmidutz, T. F., Gotlibovych, I., Smith, R. P. \&
  Hadzibabic, Z. Bose-Einstein condensation of atoms in a uniform potential.
  {\it Phys. Rev. Lett.} {\bf 110}, 200406 (2013).}

\bibitem[{Hom({\natexlab{a}})}]{Homog1}
\bibinfo{note}{Chomaz, L., Corman, L., Bienaim\'{e}, T., Desbuquois, R.,
  Weitenberg, C., Nascimb\`{e}ne, S., Beugnon, J. \& Dalibard, J. Emergence of
  coherence via transverse condensation in a uniform quasi-two-dimensional Bose
  gas. {\it Nat. Commun.} {\bf 6}, 6162 (2015).}

\bibitem[{Hom({\natexlab{b}})}]{Homog2}
\bibinfo{note}{Mukherjee, B., Yan, Z., Patel, P. B., Hadzibabic, Z., Yefsah,
  T., Struck, J. \& Zwierlein, M. W. Homogeneous atomic Fermi gases. {\it Phys.
  Rev. Lett.} {\bf 118}, 123401 (2017).}

\bibitem[{Hom({\natexlab{c}})}]{Homog3}
\bibinfo{note}{Hueck, K., Luick, N., Sobirey, L., Siegl, J., Lompe, T. \&
  Moritz, H. Two-dimensional homogeneous Fermi gases. {\it Phys. Rev. Lett.}
  {\bf 120}, 060402 (2018).}

\bibitem[{Hom({\natexlab{d}})}]{Homog4}
\bibinfo{note}{Garratt, S. J., Eigen, C., Zhang, J., Turz\'{a}k, P., Lopes, R.,
  Smith, R. P., Hadzibabic, Z. \& Navon, N. From single-particle excitations to
  sound waves in a box-trapped atomic Bose-Einstein condensate. {\it Phys. Rev.
  A} {\bf 99}, 021601(R) (2019).}

\bibitem[{Kru()}]{KrutitskyPhysRep2015}
\bibinfo{note}{Krutitsky, K. V. Ultracold bosons with short-range interaction
  in regular optical lattices. {\it Phys. Rep.} {\bf 607}, 1 (2016).}

\bibitem[{Mon()}]{MonienPRB1996}
\bibinfo{note}{Freericks, J. K. \& Monien, H. Strong-coupling expansions for
  the pure and disordered Bose-Hubbard model. {\it Phys. Rev. B} {\bf 53}, 2691
  (1996).}

\bibitem[{Eck()}]{EckardtPRB2009}
\bibinfo{note}{Teichmann, N., Hinrichs, D., Holthaus, M. \& Eckardt, A.
  Process-chain approach to the Bose-Hubbard model: Ground-state properties and
  phase diagram. {\it Phys. Rev. B} {\bf 79}, 224515 (2009).}

\bibitem[{San()}]{SansonePRB2007}
\bibinfo{note}{Capogrosso-Sansone, B., Prokof'ev, N. V. \& Svistunov, B. V.
  Phase diagram and thermodynamics of the three-dimensional Bose-Hubbard model.
  {\it Phys. Rev. B} {\bf 75}, 134302 (2007).}

\bibitem[{Dup()}]{DupuisPRB2011}
\bibinfo{note}{Ran\ifmmode\mbox{\c{c}}\else \c{c}\fi{}on, A. \& Dupuis, N.
  Nonperturbative renormalization group approach to strongly correlated lattice
  bosons. {\it Phys. Rev. B} {\bf 84}, 174513 (2011).}

\bibitem[{Sen()}]{SenguptaPRB2012}
\bibinfo{note}{Dutta, A., Trefzger, C. \& Sengupta, K. Projection operator
  approach to the Bose-Hubbard model. {\it Phys. Rev. B} {\bf 86}, 085140
  (2012).}

\bibitem[{Ger()}]{GerbierPRL2005}
\bibinfo{note}{Gerbier, F., Widera, A., F\"olling, S., Mandel, O., Gericke, T.
  \& Bloch, I. Phase coherence of an atomic Mott insulator. {\it Phys. Rev.
  Lett.} {\bf 95}, 050404 (2005).}

\bibitem[{Ket()}]{KetterlePRL2007}
\bibinfo{note}{Mun, J., Medley, P., Campbell, G. K., Marcassa, L. G.,
  Pritchard, D. E. \& Ketterle, W. Phase diagram for a Bose-Einstein condensate
  moving in an optical lattice. {\it Phys. Rev. Lett.} {\bf 99}, 150604
  (2007).}

\bibitem[{Gre({\natexlab{b}})}]{GreinerPhD}
\bibinfo{note}{Greiner, M. (2003). Ultracold quantum gases in three-dimensional
  optical lattice potentials. {\it PhD Thesis}.
  Ludwig-Maximilians-Universit\"at, M\"unchen. Online at
  http://edoc.ub.uni-muenchen.de/archive/00000968/.}

\bibitem[{Jul()}]{JuliennePRL1997}
\bibinfo{note}{Julienne, P. S., Mies, F. H., Tiesinga, E. \& Williams, C. J.
  Collisional stability of double Bose condensates. {\it Phys. Rev. Lett.} {\bf
  78}, 1880 (1997).}

\bibitem[{sqr()}]{sqrt3}
\bibinfo{note}{NIST/SEMATECH e-Handbook of Statistical Methods,
  http://www.itl.nist.gov/div898/handbook/.}

\bibitem[{Sto()}]{StoofPRA2001}
\bibinfo{note}{{v}an Oosten, D., van der Straten, P. \& Stoof, H. T. C. Quantum
  phases in an optical lattice. {\it Phys. Rev. A} {\bf 63}, 053601 (2001).}

\bibitem[{Kub()}]{KubaRepProg2015}
\bibinfo{note}{Dutta, O., Gajda, M., Hauke, P., Lewenstein, M., L\"uhmann,
  D.-S., Malomed, B. A., Sowi\'nski, T. \& Zakrzewski, J. Non-standard Hubbard
  models in optical lattices: a review. {\it Rep. Prog. Phys.} {\bf 78}, 066001
  (2015).}

\bibitem[{Vic({\natexlab{b}})}]{VicariPhysRep2002}
\bibinfo{note}{Pelissetto, A. \& Vicari, E. Critical phenomena and
  renormalization-group theory. {\it Phys. Rep.} {\bf 368}, 549 (2002).}

\bibitem[{Tak()}]{TakahashiNatComm2016}
\bibinfo{note}{Kato, S., Inaba, K., Sugawa, S., Shibata, K., Yamamoto, R.,
  Yamashita, M. \& Takahashi, Y. Laser spectroscopic probing of coexisting
  superfluid and insulating states of an atomic Bose-Hubbard system. {\it Nat.
  Commun.} {\bf 7}, 11341 (2016).}

\bibitem[{3DB()}]{3DBHexp}
\bibinfo{note}{Nakamura, Y., Takasu, Y., Kobayashi, J., Asaka, H., Fukushima,
  Y., Inaba, K., Yamashita, M. \& Takahashi, Y. Experimental determination of
  Bose-Hubbard energies. {\it Phys. Rev. A} {\bf 99}, 033609 (2019).}

\bibitem[{exp()}]{expNext}
\bibinfo{note}{Zhou, T., Yang, K., Zhu, Z., Yu, X., Yang, S., Xiong, W., Zhou,
  X., Chen, X., Li, C., Schmiedmayer, J., Yue, X. \& Zhai, Y. Observation of
  atom-number fluctuations in optical lattices via quantum collapse and revival
  dynamics. {\it Phys. Rev. A} {\bf 99}, 013602 (2019).}

\bibitem[{Pol()}]{PolletRepProgPhys2012}
\bibinfo{note}{Pollet, L. Recent developments in quantum Monte Carlo
  simulations with applications for cold gases. {\it Rep. Prog. Phys.} {\bf
  75}, 094501 (2012).}

\bibitem[{Yba()}]{YbaPRA2017}
\bibinfo{note}{Borkowski, M., Buchachenko, A. A., Ciury\l{}o, R., Julienne, P.
  S., Yamada, H., Kikuchi, Y., Takahashi, K., Takasu, Y. \& Takahashi, Y.
  Beyond-Born-Oppenheimer effects in sub-kHz-precision photoassociation
  spectroscopy of ytterbium atoms. {\it Phys. Rev. A} {\bf 96}, 063405 (2017).}

\bibitem[{Pad()}]{PadeBook}
\bibinfo{note}{Baker, G. A. \& Graves-Morris, P. {\it Pad\'{e} Approximants}
  (Cambridge University Press, 2nd edition, 1996).}

\bibitem[{Car({\natexlab{b}})}]{Cardy1}
\bibinfo{note}{Cardy, J. L., ed., {\it Finite-Size Scaling} (North-Holland,
  Amsterdam, 1988).}

\bibitem[{Mat()}]{Mathematica}
\bibinfo{note}{Wolfram Research, Inc., Mathematica, Version 11.0, Champaign, IL
  (2016).}

\bibitem[{Bax()}]{Baxter}
\bibinfo{note}{Baxter, R. J. {\it Exactly Solved Models in Statistical
  Mechanics} (Academic Press, London, 1982).}

\bibitem[{Lie()}]{Lieb1961}
\bibinfo{note}{Lieb, E., Schultz, T. \& Mattis, D. Two soluble models of an
  antiferromagnetic chain. {\it Ann. Phys. (N.Y.)} {\bf 16}, 407 (1961).}

\bibitem[{Pfe()}]{PfeutyAnn1970}
\bibinfo{note}{Pfeuty, P. The one-dimensional Ising model with a transverse
  field. {\it Ann. Phys.} {\bf 57}, 79 (1970).}

\bibitem[{Fra()}]{FranzPRL2006}
\bibinfo{note}{Franz, M. \& Iyengar, A. P. Superfluid density of strongly
  underdoped cuprate superconductors from a four-dimensional $XY$ model. {\it
  Phys. Rev. Lett.} {\bf 96}, 047007 (2006).}

\bibitem[{Bro()}]{BrounPRL2007}
\bibinfo{note}{Broun, D. M., Huttema, W. A., Turner, P. J., \"Ozcan, S.,
  Morgan, B., Liang, R., Hardy, W. N. \& Bonn, D. A. Superfluid density in a
  highly underdoped ${\mathrm{YBa}}_{2}{\mathrm{Cu}}_{3}{\mathrm{O}}_{6+y}$
  superconductor. {\it Phys. Rev. Lett.} {\bf 99}, 237003 (2007).}

\bibitem[{Egg()}]{EggelPRB2011}
\bibinfo{note}{Eggel, T., Oshikawa, M. \& Shirahama, K. Four-dimensional $XY$
  quantum critical behavior of ${}^{4}$He in nanoporous media. {\it Phys. Rev.
  B} {\bf 84}, 020515(R) (2011).}

\bibitem[{alp({\natexlab{a}})}]{alps1}
\bibinfo{note}{Albuquerque, A. F. {\it et al.} The ALPS project release 1.3:
  Open-source software for strongly correlated systems. {\it J. Magn. Magn.
  Matter.} {\bf 310}, 1187 (2007).}

\bibitem[{alp({\natexlab{b}})}]{alps2}
\bibinfo{note}{Bauer, B. {\it et al.} The ALPS project release 2.0: open source
  software for strongly correlated systems. {\it J. Stat. Mech.} P05001
  (2011).}

\end{thebibliography}

\end{document}